\documentclass[aps,prc,groupedaddress,showkeys,showpacs,manuscript]{revtex4}
\usepackage{amssymb}
\usepackage{array}
\usepackage{graphicx}
\usepackage{color}
\usepackage{CJKutf8}
\newcommand{\upcite}[1]{\textsuperscript{\textsuperscript{\cite{#1}}}}

\begin{document}
\begin{CJK}{UTF8}{gbsn}

\title{Transverse Momentum Distribution of kaons, pions, and (anti-)protons production in U+U collisions at $\sqrt{s_{NN}}$ = 193 GeV using the UrQMD Model}

\author {Ying Yuan,$\, ^{1, 2}$\footnote{E-mail address: yuany@gxtcmu.edu.cn}}
\address{
1) College of Pharmacy, Guangxi University of Chinese Medicine, Nanning 530200, China\\
2) Guangxi Key Laboratory of Nuclear Physics and Nuclear Technology, Guangxi Normal University, Guilin 541004, China\\}


\begin{abstract}
In this study, we investigate the transverse momentum spectra of $K ^{\pm }$, $\pi ^{\pm }$ and $p(\bar{p})$ in mid-rapidity ($\left | y \right | < 0.1$) for nine centrality classes ranging from $0\%$ to $80\%$ in $^{238} U$+$^{238} U$ collisions at $\sqrt{s_{NN}}$=193 GeV. The simulations are performed using the Ultra-relativistic Quantum Molecular Dynamics (UrQMD) model, specifically employing both the cascade mode and the soft momentum-dependent equation of state (SM-EoS) mode. Additionally, we extract other observables from the $p_{T}$ spectrum, including the average transverse momentum ($\left \langle p_{T}  \right \rangle$), the particle yield ($dN/dy$) and particle-type ratios, presenting them as a function of collision centrality. We find that the collision dynamics are significantly sensitive to the prolate deformation of the uranium nuclei, which influences the initial geometry and subsequent particle production. We find that the U+U collision dynamics are highly sensitive to the deformation of the uranium nucleus. Consequently, the cascade mode is more appropriate for describing the low-$p_{T}$ region ($p_{T} < 1.2 GeV/c$), while the SM-EoS mode better captures the trends in the high-$p_{T}$ region ($p_{T} > 1.2 GeV/c$). Furthermore, at RHIC energies, our results indicate that pair production is the dominant mechanism for particle creation in the mid-rapidity region. This conclusion is corroborated by the particle-to-antiparticle production ratio, which approaches unity—indicating a high degree of matter–antimatter symmetry in the observed collision events.

\end{abstract}



\maketitle

{\section{Introduction}}

Heavy-ion collisions (HICs) at ultra-relativistic energies provide a unique laboratory for exploring the properties of strongly interacting matter under extreme conditions of temperature and density \upcite{Alt1,Sun2,Wang3,Li4,Liu5,Abdulhamid6,Adam7}. A primary goal of these experiments is to understand the phase transition of quantum chromodynamics (QCD) from the quark-gluon plasma (QGP) to the hadron gas phase \upcite{Arsen8,Li9}. Over the past two decades, significant progress has been made at the Relativistic Heavy Ion Collider (RHIC), particularly in mapping the QCD phase diagram near the critical temperature \upcite{Lao10}. Theoretical frameworks such as statistical models, coalescence models, and microscopic transport models have been extensively employed to describe particle production and evolution \upcite{Mrowc11,Mrowc12,Bazak13,Liu14,Liu15,Yuan16,Yuan17,Peng18,Yuan19,Yue20,Liu21}. Among these, transport models are particularly crucial for understanding the dynamical evolution of the system, as transverse momentum spectra offer critical insights into the freeze-out conditions and the collective dynamics of the interacting matter \upcite{Li22,Chen23}.
 
While symmetric systems like Au+Au have been widely studied, U+U collisions offer a distinct advantage due to the non-spherical shape of the uranium nucleus. Unlike spherical nuclei, $^{238} U$ nuclei possess a significant prolate quadrupole deformation. This deformation introduces a strong dependence of the initial collision geometry on the relative orientation of the colliding nuclei, thereby providing a sensitive probe for the initial state geometry and its impact on final-state observables such as elliptic flow ($v_{2}$) and particle yields \upcite{STAR24}.

Recent experimental results from the STAR Collaboration have confirmed that the nuclear deformation significantly influences the collective flow and particle production \upcite{Abdallah25,STAR26}. Specifically, recent measurements of charged particle multiplicities and elliptic flow in U+U collisions have shown a clear sensitivity to the nuclear deformation parameter, validating the importance of initial geometry effects in heavy-ion collisions \upcite{STAR24}. Furthermore, recent studies on identified particle spectra in U+U collisions have provided new constraints on the equation of state and the chemical freeze-out conditions \upcite{Abdallah25}. 

In this study, we utilize the Ultra-Relativistic Quantum Molecular Dynamics (UrQMD) transport model to investigate the transverse momentum distributions of $\pi$(pions) mesons, $K$(kaons), and $p$($\bar{p}$)(protons/antiprotons) in U+U collisions at $\sqrt{s_{NN}}$=193 GeV. We specifically examine the role of collisional interactions and compare our theoretical results with recent experimental data from the STAR Collaboration \upcite{Abdallah25}. The primary objective is to elucidate the mechanisms of particle production in deformed nuclear collisions and to assess the capability of transport models in describing the final-state particle spectra in U+U systems.

\vspace{1\baselineskip}

{\section{Ultrarelativistic quantum molecular dynamics transport model}}

{\subsection{The ultra-relativistic quantum molecular dynamics (UrQMD) model}}

The UrQMD model is a microscopic many-body approach to transport that can be applied to study the interactions of protons with protons (pp), with protons on a nucleus (pA) with a nucleus on a nucleus (AA) over the energy range from the Spheron Ion Source to the Large Hadron Collider. This model is based on the propagation of color strings, of which the constituent quarks and diquarks (as the end of the string) carry mesonic and baryonic degrees of freedom \upcite{Petersen27}. It can combine different reaction mechanisms and provide theoretical simulations of various experimental observations. Currently, in our model, subhadrons' degrees of freedom enter through a formation time for hadrons resulting from string fragmentation, which is dominant at the early stages of heavy-ion collisions (HICs) at high SPS and RHIC energies \upcite{Andersson28,Nilsson29,Sjostrand30}.

\vspace{1\baselineskip}

Like the quantum molecular dynamics (QMD) model, the UrQMD model follows parallel principles to simulate hadrons in phase space with the propagation of each individual hadron's phase space according to Hamilton's equation of motion \upcite{Bass31},
\begin{equation}
\dot{\vec{r_{i}}}=\frac{\partial{H}}{\partial{\vec{p_{i}}}},
\hspace{1cm}
\dot{\vec{p_{i}}}=-\frac{\partial{H}}{\partial{\vec{r_{i}}}}.
\label{eq:1}
\end{equation}
Here, $\vec{r_{i}}$ and $\vec{p_{i}}$ are the coordinate and momentum of the hadron \emph{i}, and the Hamiltonian H consists of the kinetic energy T and the effective interaction potential energy U,
\begin{equation}
H=T+U.  \label{eq:2}
\end{equation}

This microscopic transport approach simulates the interactions of incoming and newly produced particles, the excitation and fragmentation of color strings and the formation and decay of hadronic resonances. In the pursuit of higher energies, it is crucial to consider the treatment of subhadronic degrees of freedom. In this current version, the degrees of freedom are introduced through a hadron formation time, in the string fragmentation process, and there is no explicit incorporation of a phase transition from a hadronic to a quark-gluon phase in the model's dynamics. However, a detailed analysis of the model in thermal equilibrium yields an effective equation of state of the Hagedorn type \upcite{Li32}.

For the initialization part, the nucleon density distribution in the ground state follows the Woods Saxon distribution,
\begin{equation}
\rho(r,\theta ,\phi)=\frac{\rho _{0}}{1+e^{[(r-R(\theta ,\phi )) /a]}}
\end{equation}
Here, $\rho _{0}$=0.16 $fm^{-3} $, $a=0.55 fm$. Considering the quadrupole deformation of uranium nuclei \upcite{Gao33},
\begin{equation}
R(\theta ,\phi )=R_{0}[1+\beta _{2}Y_{20}(\theta ,\phi )]
\end{equation}
and $R_{0}=1.16A^{\frac{1}{3}}$, $\beta _{2}=0.28$ \upcite{Ma34,Zhao35}. In the calculation of this paper, we simulate the results of the random orientation collision between the projectile and the target core.

{\subsection{The soft momentum dependent equation of state (SM-EoS)}}

In the standard framework of the UrQMD model, the term "potential energy" incorporates various types of interactions such as those involving the two-body and three-body Skyrme-, Yukawa-, Coulomb- and Pauli-terms \upcite{Yuan16,Bass31,Bleicher36},
\begin{equation}
U=U_{\rm sky}^{(2)}+U_{\rm sky}^{(3)}+U_{\rm Yuk}+U_{\rm Cou}+U_{\rm pau}.
\label{eq:3}
\end{equation}

In the upgraded UrQMD (version $3.4$) of the present work, additional terms are defined: (1) the density-dependent symmetry potential term $U_{\rm sym}$ and (2) the momentum-dependent term $U_{\rm md}$ \upcite{Bass37}. In this study, the soft momentum-dependent (SM) equation of state (EoS) is applied, which is presented in Ref. \cite{Li38}. In the RHIC energy regime, the Yukawa-, Pauli-, and baryon symmetry potentials become unimportant, while the Skyrme and momentum-dependence parts of potentials still affect the full HIC dynamic process \upcite{Li39}. During the formation time, the "pre-formed" particles (string fragments that will be abducted onto hadronic states later on) are usually treated to be free streaming, while reduced cross sections are only included for leading hadrons. 

In this paper, the transverse momentum distributions and central yields of $\pi$ mesons, $K$ mesons and $p$($\bar{p}$) generated in U+U collisions at $\sqrt{s_{NN}}$=193 GeV at mid-rapidity ($\vert{y}\vert$$<$0.1) are studied by using UrQMD cascade mode and SM-EoS mode.

\vspace{1\baselineskip}

{\section{Results and discussions}}

{\subsection{Transverse momentum spectrum}}

\begin{figure}
\setlength{\abovedisplayskip}{-0.5cm}
\includegraphics[angle=0,width=17.6cm]{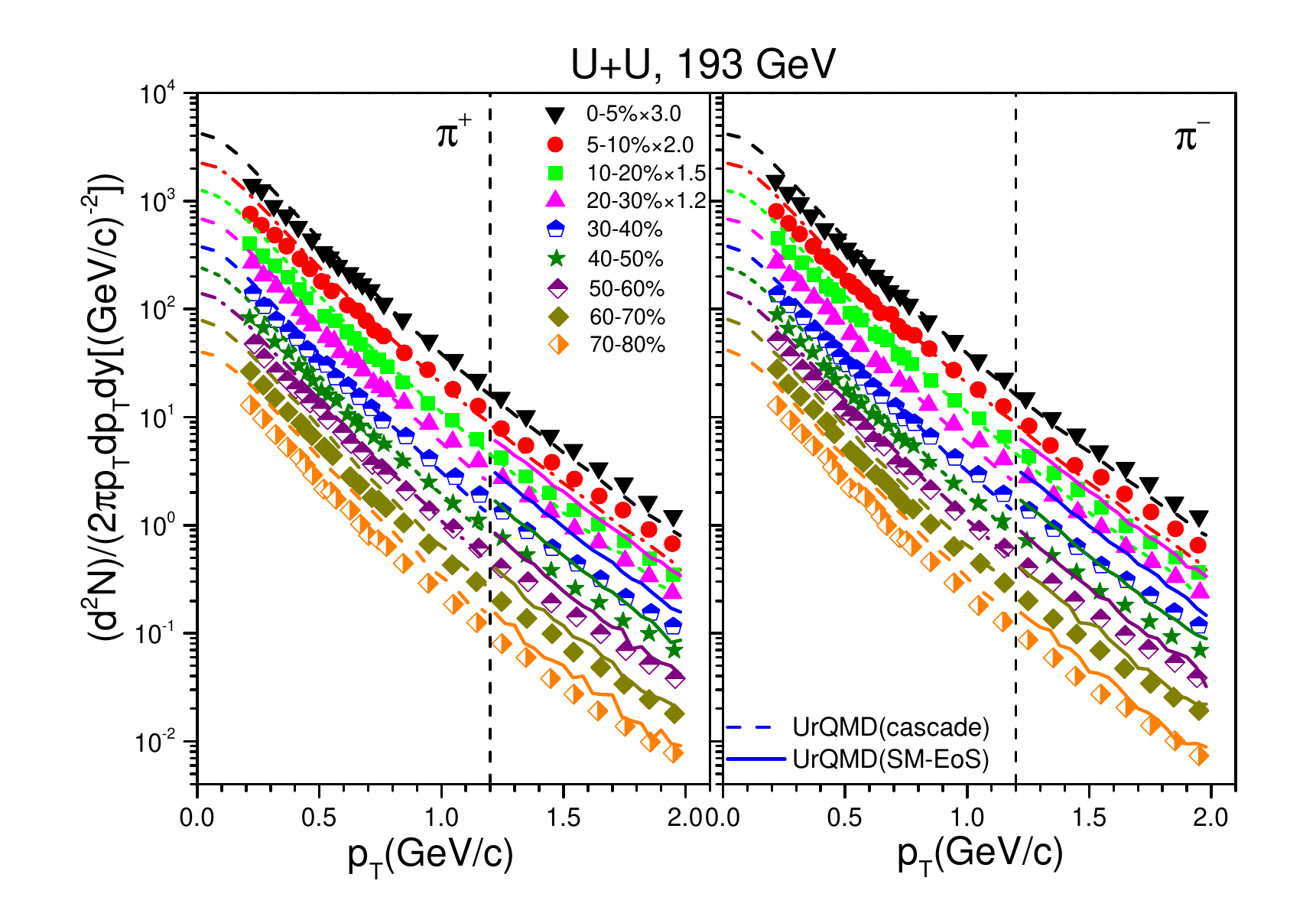}
\caption{Transverse momentum spectra of $\pi^{+}$ and $\pi^{-}$ are calculated at mid-rapidity ($\left | y \right | < 0.1$) in U+U collisions at $\sqrt{s_{NN}}$=193 GeV for $0-5\%$, $5-10\%$, $10-20\%$, $20-30\%$, $30-40\%$, $40-50\%$, $50-60\%$, $60-70\%$ and $70-80\%$ centralities from the cascade mode and the soft momentum dependent equation of state mode of UrQMD model. The lines denote calculations, while the symbol represents experimental data taken from the STAR Collaboration \upcite{Abdallah25}.} \label{fig1}
\end{figure}

Fig.\ \ref{fig1} is shown the transverse momeutum spectra in nine centrality classes in U+U collisions at $\sqrt{s_{NN}}$=193 GeV at mid-rapidity ($\vert{y}\vert$$<$0.1) for $\pi^{+}$ and $\pi^{-}$. There are nine centrality classes, representing a range of $0-5\%$, $5-10\%$, $10-20\%$, $20-30\%$, $30-40\%$, $40-50\%$, $50-60\%$, $60-70\%$ and $70-80\%$ respectively. The dotted lines are the results calculated from the cascade mode of UrQMD model. The solid lines are the results calculated from the soft momentum dependent equation of state mode of UrQMD model. The symbols are the experimental data from the STAR Collaboration \upcite{Abdallah25}. The calculations are shown for $p_{T} < 2.0 GeV/c$ in the Figure. The cascading mode of the UrQMD model has been observed to demonstrate significant agreement with empirical laws at centralities of $20-80\%$ for $p_{T} < 1.2 GeV/c$ and at $0-20\%$ centrality. However, in the transverse momentum region at $20-80\%$ with $1.2 GeV/c < p_{T} < 2.0 GeV/c$, the theoretical yield of the SM-EoS model exceeds the experimental measurements. However, the higher the momentum is, the closer the theoretical description is to the experimental value. Due to the limitations of the UrQMD model in the physical description of some partons, it overestimates the yield of $\pi$ mesons in the medium transverse momentum region.

\begin{figure}
\setlength{\abovecaptionskip}{-0.5cm}
\includegraphics[angle=0,width=17.6cm]{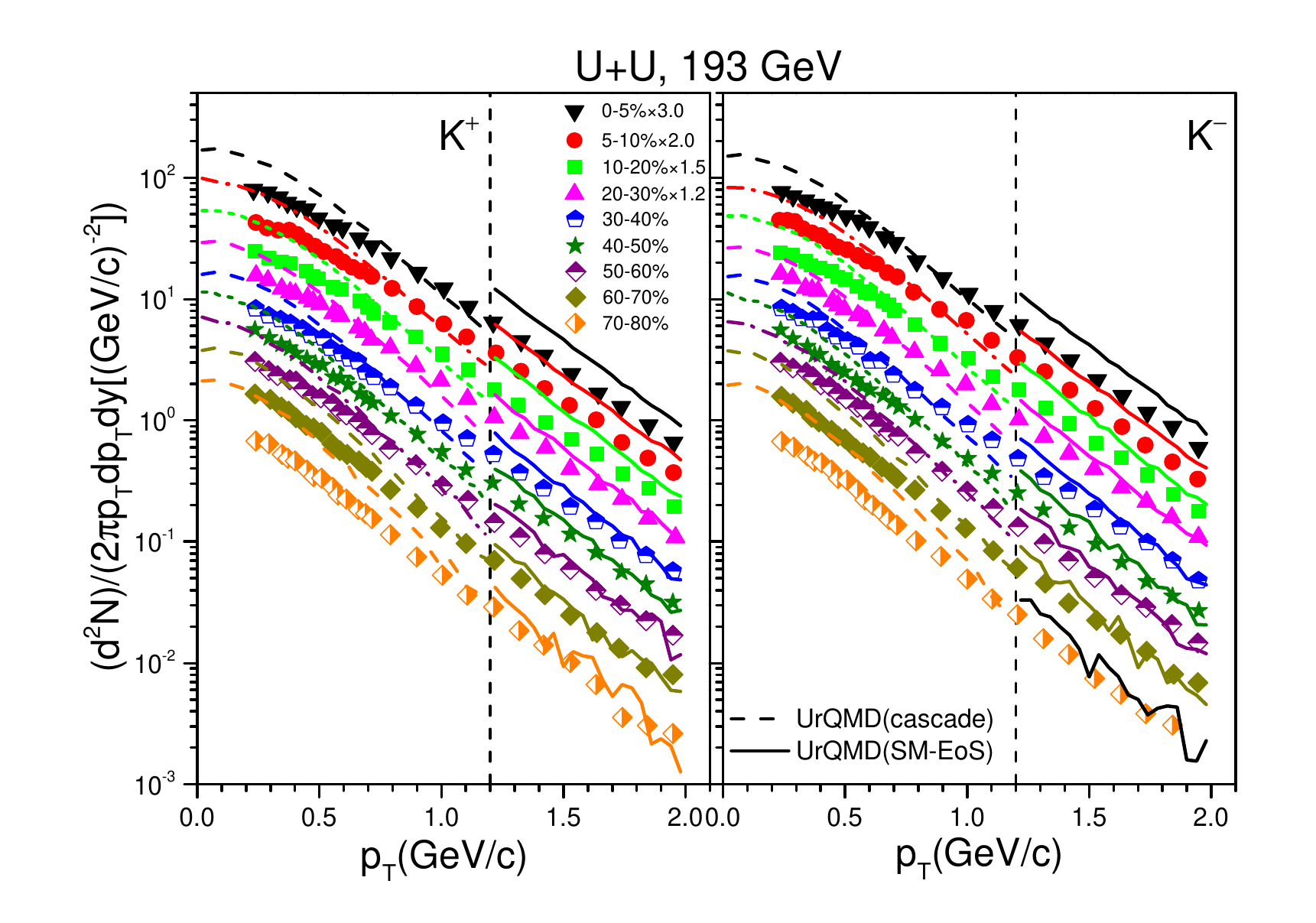}
\caption{Transverse momentum spectra of $K^{+}$ and $K^{-}$ are calculated at mid-rapidity ($\left | y \right | < 0.1$) in U+U collisions at $\sqrt{s_{NN}}$=193 GeV for $0-5\%$, $5-10\%$, $10-20\%$, $20-30\%$, $30-40\%$, $40-50\%$, $50-60\%$, $60-70\%$ and $70-80\%$ centralities from the cascade mode and the soft momentum dependent equation of state mode of UrQMD model. The lines denote calculations, while the symbol represents experimental data taken from the STAR Collaboration \upcite{Abdallah25}.} \label{fig2}
\end{figure}

Fig.\ \ref{fig2} is shown the transverse momeutum spectra in nine centrality classes in U+U collisions at $\sqrt{s_{NN}}$=193 GeV at mid-rapidity ($\vert{y}\vert$$<$0.1) for $K^{+}$ and $K^{-}$. The dotted lines are the results calculated from the cascade mode of UrQMD model. The solid lines are the results calculated from the soft momentum dependent equation of state mode of UrQMD model. The symbols are the experimental data from the STAR Collaboration \upcite{Abdallah25}. The calculations for transverse momentum ($p_{T}$) below 2.0 GeV/c are presented in the figure. The cascading mode of the UrQMD model has been found to exhibit significant agreement with empirical laws for $p_{T} < 1.2$ GeV/c. However, it is observed that as the transverse momentum decreases, the theoretical values tend to be overestimated. Furthermore, with an increase in collision centrality, the range of overestimation in transverse momentum also expands. In the region where $1.2 GeV/c < p_{T} < 2.0 GeV/c$ at centralities ranging from $20\%$ to $80\%$, the theoretical predictions align closely with experimental results. Conversely, at a centrality of $0-20\%$, the theoretical yield predicted by the SM-EoS model surpasses experimental measurements. This difference can be attributed to the fact that the UrQMD model cannot directly explain the phase transition effect in the quark-gluon plasma (QGP). Furthermore, the influence of the quark recombination mechanism on the yield of $k$ mesons has not been considered either. Consequently, within central collision regions, there is an overestimation of transverse momentum distributions for $K$ mesons.

\begin{figure}
\setlength{\abovecaptionskip}{-0.5cm}
\includegraphics[angle=0,width=17.6cm]{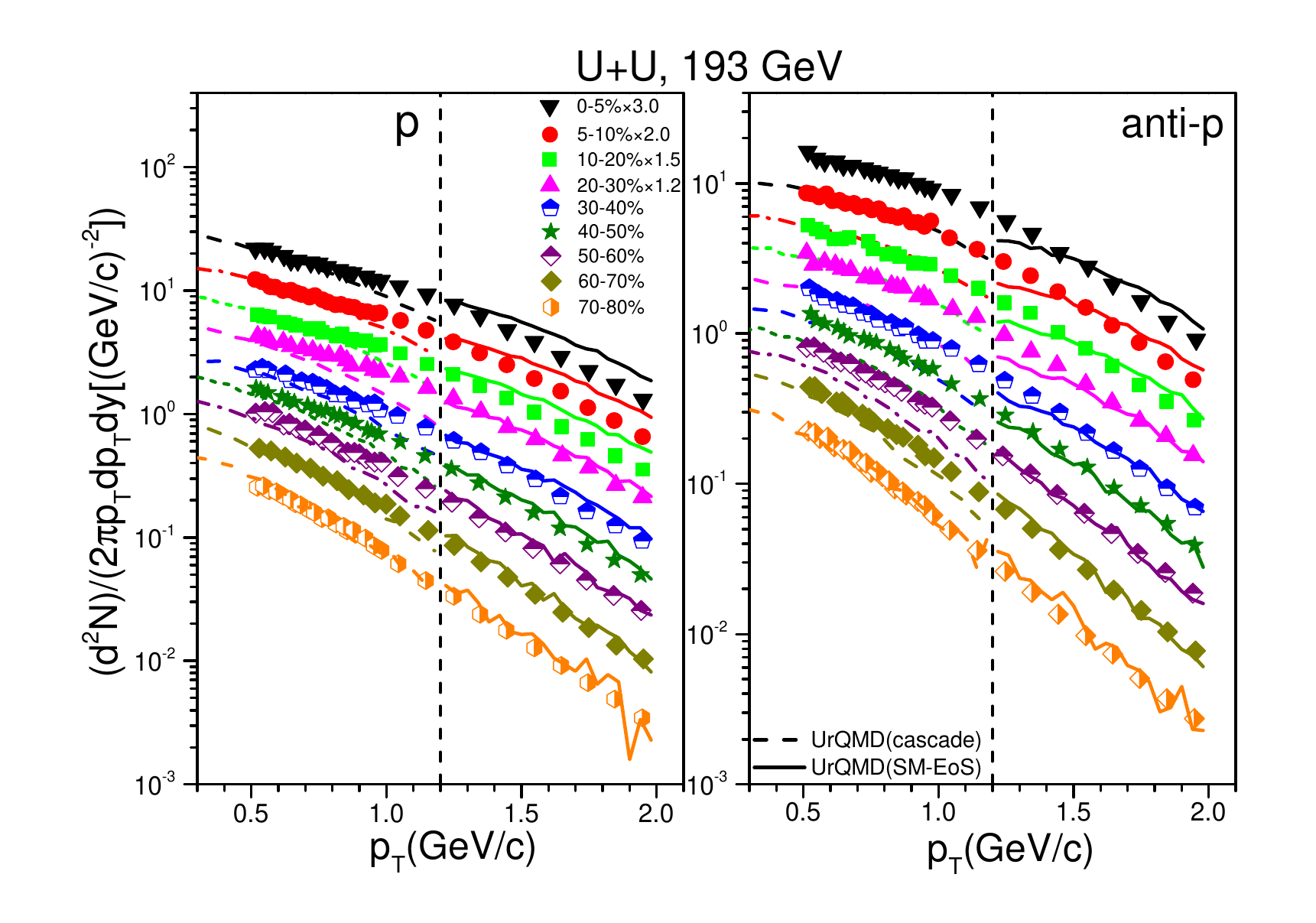}
\caption{Transverse momentum spectra of $p$ and $\bar{p}$ are calculated at mid-rapidity ($\left | y \right | < 0.1$) in U+U collisions at $\sqrt{s_{NN}}$=193 GeV for $0-5\%$, $5-10\%$, $10-20\%$, $20-30\%$, $30-40\%$, $40-50\%$, $50-60\%$, $60-70\%$ and $70-80\%$ centralities from the cascade mode and the soft momentum dependent equation of state mode of UrQMD model. The lines denote calculations, while the symbol represents experimental data taken from the STAR Collaboration \upcite{Abdallah25}.} \label{fig3}
\end{figure}

Fig.\ \ref{fig3} presents the transverse momentum spectra across nine centrality classes in U+U collisions at $\sqrt{s_{NN}}$=193 GeV, measured at mid-rapidity ($\vert{y}\vert$$<$0.1) for protons ($p$) and antiprotons ($\bar{p}$). The dotted lines represent results calculated using the cascade mode of the UrQMD model, while the solid lines correspond to outcomes derived from a soft momentum-dependent equation of state within the same model. The symbols denote experimental data obtained from the STAR Collaboration \upcite{Abdallah25}. The calculations are displayed for $p_{T} < 2.0 GeV/c$ in this figure. It is evident from the figure that for protons for $p_{T} < 2.0 GeV/c$ and antiprotons for $p_{T} > 1.2 GeV/c$, the theoretical predictions of transverse momentum distributions align well with experimental observations. However, in the region where $p_{T} < 1.2 GeV/c$, there is a noticeable underestimation of theoretical values for antiprotons as collision centrality decreases. This discrepancy arises because the UrQMD model does not account for quark-gluon plasma phase transitions, which may lead to an underprediction of antiproton yields in high-temperature and high-density environments.

Our analysis reveals a distinct behavior of the UrQMD model across different transverse momentum regions. In the low-$p_{T}$ region ($p_{T} < 1.2 GeV/c$), the cascade mode provides a superior description of the experimental data. Conversely, in the intermediate-to-high $p_{T}$ region ($1.2 GeV/c < p_{T} < 2.0 GeV/c$), the SM-EoS mode yields a better fit to the data. This improvement suggests that mean-field interactions (or the soft equation of state effects) play a significant role in modifying the particle spectra at higher momenta, likely through their influence on the scattering cross-sections and the expansion dynamics. To understand this transition, we consider the prolate deformation of the $^{238} U$ nuclei. In our simulations, we account for the quadrupole deformation and assume an equal probability for sharp-tip, body-body, and body-tip collision geometries. The time evolution of the collision process further elucidates these findings: during the early stage, high energy density in the central region leads to significant re-scattering, where the system's behavior is largely governed by local collision dynamics rather than global mean fields. As the system evolves, particles propagate outward, and the cumulative effect of the mean field becomes increasingly pronounced, thereby enhancing the performance of the SM-EoS mode in the higher $p_{T}$ region.

\vspace{1\baselineskip}

{\subsection{Average transverse momentum distributions}}

Figure 4 shows the variation of $\left \langle p_{T}  \right \rangle$ with $\left \langle N_{part}  \right \rangle$ at midrapidity ($\left | y \right | < 0.1$) for $\pi^{+}$, $K^{+}$ and $p$ particles in U+U collisions at $\sqrt{s_{NN}}$=193 GeV. The black diamonds were obtained from UrQMD calculations, and the red solid circles are the experimental data \upcite{Abdallah25}. It is found that the experimental results can be described within the tolerance of error. The values of $\left \langle p_{T}  \right \rangle$ increase slowly with the decrease of collision centrality, and they are listed in Table I.

\begin{figure}[htbp] \centering
\setlength{\abovecaptionskip}{-1cm}
\includegraphics[angle=0,width=17.6cm]{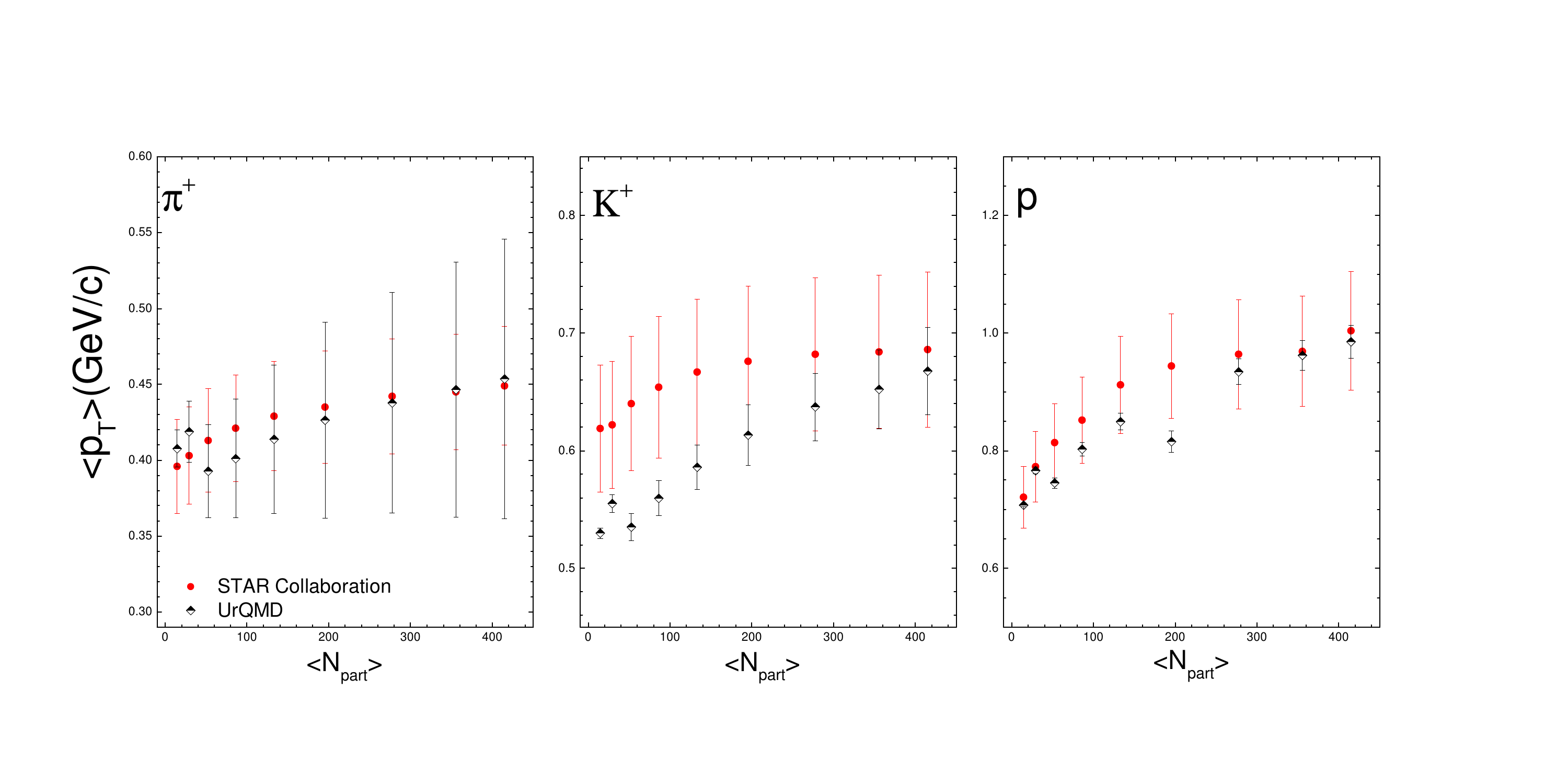}
\caption{$\left \langle p_{T}  \right \rangle$ as a function of $\left \langle N_{part}  \right \rangle$ at mid-rapidity ($\left | y \right | < 0.1$) of $\pi^{+}$, $K^{+}$ and $p$ for U+U collisions at $\sqrt{s_{NN}}$=193 GeV. The red solid circles represent data collected by the STAR Collaboration \upcite{Abdallah25}. The black diamonds are the calculation using UrQMD model. } \label{fig4}
\end{figure}

\begin{table}
\caption{Values of $\left \langle p_{T}  \right \rangle$ in GeV/c within mid-rapidity ($\left | y \right | < 0.1$) of $\pi^{+}$, $\pi^{-}$, $K^{+}$, $K^{-}$, $p$ and $\bar{p}$ for U+U collisions at $\sqrt{s_{NN}}$=193 GeV using the UrQMD model.} \label{TableI}
\begin{tabular}{ p{1.8cm}<{\centering} p{2.3cm}<{\centering} p{2.3cm}<{\centering} p{2.3cm}<{\centering} p{2.3cm}<{\centering} p{2.3cm}<{\centering} p{2.3cm}<{\centering}}
\hline \hline
      Centrality & $\pi^{+}$ & $\pi^{-}$ & $K^{+}$ & $K^{-}$ & $p$ & $\bar{p}$ \\
\hline
      0-5\% & $0.450\pm 0.092$ & $0.450\pm 0.092$ & $0.661\pm0.037$ & $0.661\pm0.035$ & $0.980\pm0.028$ & $1.053\pm0.019$ \\
      5-10\% & $0.445\pm 0.082$ & $0.445\pm 0.082$ & $0.651\pm0.033$ & $0.649\pm0.031$ & $0.960\pm0.025$ & $1.028\pm0.017$ \\
      10-20\% & $0.440\pm 0.070$ & $0.440\pm 0.070$ & $0.638\pm0.028$ & $0.634\pm0.026$ & $0.932\pm0.021$ & $1.003\pm0.015$ \\
      20-30\% & $0.432\pm 0.057$ & $0.432\pm 0.057$ & $0.620\pm0.022$ & $0.616\pm0.021$ & $0.915\pm0.017$ & $0.961\pm0.013$ \\
      30-40\% & $0.423\pm 0.045$ & $0.424\pm 0.045$ & $0.599\pm0.018$ & $0.596\pm0.017$ & $0.885\pm0.014$ & $0.919\pm0.011$ \\
      40-50\% & $0.415\pm 0.036$ & $0.415\pm 0.036$ & $0.580\pm0.014$ & $0.574\pm0.013$ & $0.844\pm0.011$ & $0.874\pm0.009$ \\
      50-60\% & $0.408\pm 0.027$ & $0.409\pm 0.027$ & $0.563\pm0.010$ & $0.558\pm0.010$ & $0.803\pm0.008$ & $0.820\pm0.007$ \\
      60-70\% & $0.388\pm 0.026$ & $0.389\pm 0.026$ & $0.522\pm0.010$ & $0.518\pm0.009$ & $0.750\pm0.006$ & $0.880\pm0.006$ \\
      70-80\% & $0.397\pm 0.014$ & $0.400\pm 0.014$ & $0.519\pm0.005$ & $0.519\pm0.005$ & $0.695\pm0.004$ & $0.715\pm0.004$ \\
\hline \hline
\end{tabular}
\end{table}

\vspace{1\baselineskip}

{\subsection{Particle yields}}

\begin{figure}[htbp] \centering
\setlength{\abovecaptionskip}{-0.5cm}
\includegraphics[angle=0,width=17.6cm]{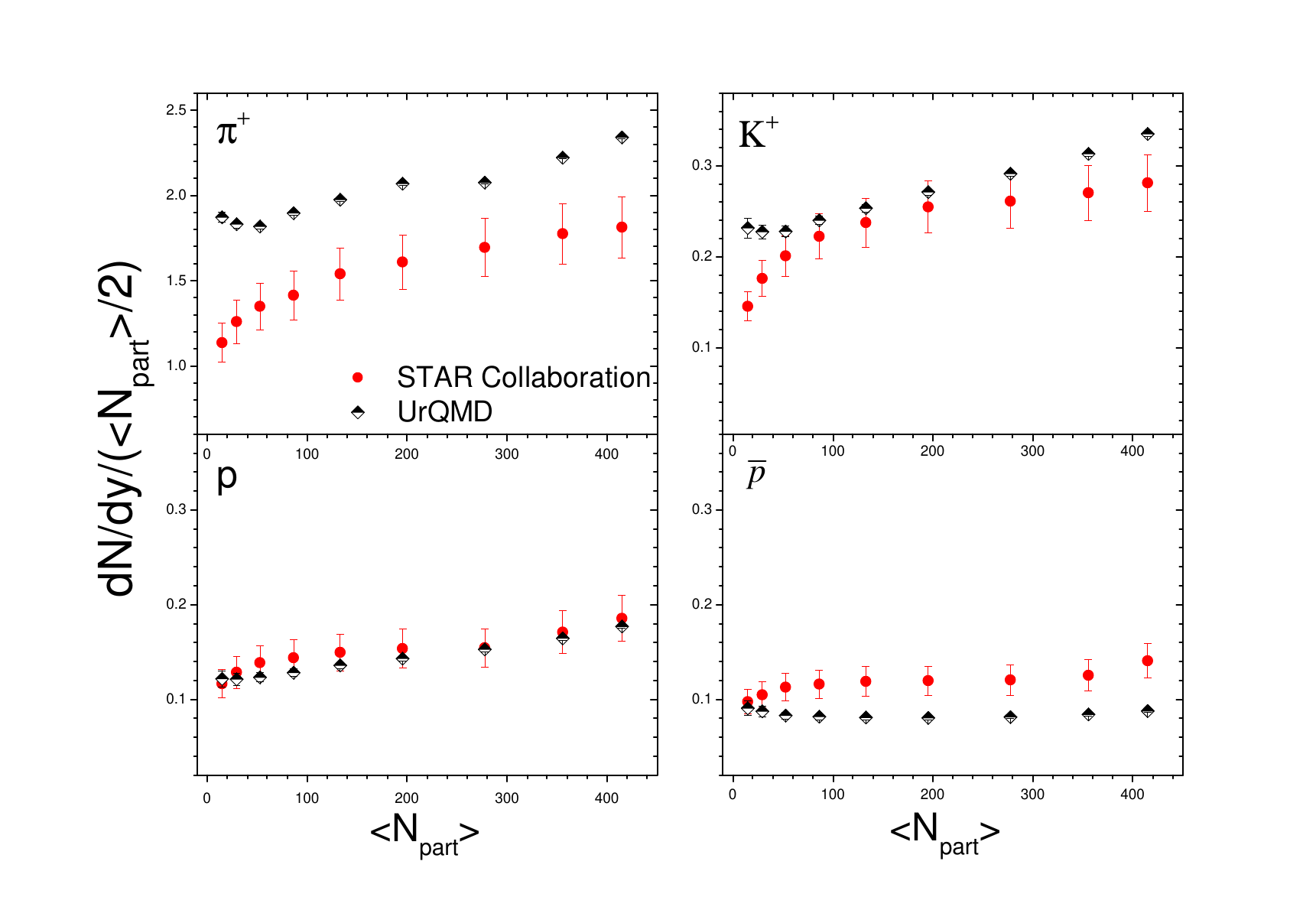}
\caption{$dN/dy$ by $\left \langle N_{part}  \right \rangle /2$ as a function of $\left \langle N_{part}  \right \rangle$ at mid-rapidity ($\left | y \right | < 0.1$) of $\pi^{+}$, $K^{+}$, $p$ and $\bar{p}$ for U+U collisions at $\sqrt{s_{NN}}$=193 GeV. The red solid circles represent experimental data taken from the STAR Collaboration \upcite{Abdallah25}. The black diamonds are the calculation using UrQMD model. } \label{fig5}
\end{figure}

Fig.\ \ref{fig5} shows $dN/dy$ as a function of $\left \langle N_{part} \right \rangle$ at mid-rapidity ($\left | y \right | < 0.1$) for $\pi^{+}$, $K^{+}$, $p$, and $\bar{p}$ in U+U collisions at $\sqrt{s_{NN}} = 193$ GeV. The black diamonds represent the results calculated from the UrQMD model, while the red solid circles correspond to the experimental data \upcite{Abdallah25}. The values of $dN/dy$ increase gradually with decreasing collision centrality and are summarized in Table II. It is evident that the theoretical predictions for $K^{+}$, $p$, and $\bar{p}$ agree well with the experimental data within the allowable error range. However, the $\pi^{+}$ mesons in peripheral collisions deviate significantly from the experimental results. This discrepancy arises because $\pi$ mesons are predominantly produced via nucleon-nucleon interactions, nucleon resonance decays, and baryon intermediate states. Both nucleon resonance states and baryon intermediate states significantly influence the yield of $\pi$ mesons. In peripheral collisions, the mean-field effect is considered, which slows down the escape of $\pi$ mesons from the collision zone, thereby enhancing their production in this region. Additionally, insufficient consideration of initial fluctuations in the model may contribute to this deviation. These aspects will be investigated more thoroughly in future studies.

\begin{table}
\caption{Values of $dN/dy$ within mid-rapidity ($\left | y \right | < 0.1$) of $\pi^{+}$, $\pi^{-}$, $K^{+}$, $K^{-}$, $p$ and $\bar{p}$ for U+U collisions at $\sqrt{s_{NN}}$=193 GeV using the UrQMD model.} \label{TableII}
\begin{tabular}{ p{1.8cm}<{\centering} p{2.3cm}<{\centering} p{2.3cm}<{\centering} p{2.3cm}<{\centering} p{2.3cm}<{\centering} p{2.3cm}<{\centering} p{2.3cm}<{\centering}}
\hline \hline
      Centrality & $\pi^{+}$ & $\pi^{-}$ & $K^{+}$ & $K^{-}$ & $p$ & $\bar{p}$ \\
\hline
      0-5\% & $485.32\pm 1.12$ & $518.46\pm 1.44$ & $69.52\pm0.56$ & $62.19\pm0.53$ & $36.74\pm0.44$ & $18.24\pm0.31$ \\
      5-10\% & $394.67\pm 1.01$ & $420.04\pm 1.29$ & $55.63\pm0.50$ & $49.54\pm0.47$ & $29.22\pm0.39$ & $14.93\pm0.28$ \\
      10-20\% & $287.96\pm 0.87$ & $305.06\pm 1.10$ & $40.43\pm0.43$ & $36.35\pm0.40$ & $21.19\pm0.33$ & $11.29\pm0.24$ \\
      20-30\% & $202.19\pm 0.89$ & $203.84\pm 0.89$ & $26.49\pm0.34$ & $23.91\pm0.33$ & $13.98\pm0.27$ & $7.87\pm0.20$ \\
      30-40\% & $131.33\pm 0.71$ & $132.93\pm 0.71$ & $16.85\pm0.27$ & $15.37\pm0.26$ & $9.04\pm0.21$ & $5.37\pm0.17$ \\
      40-50\% & $81.76\pm 0.55$ & $82.39\pm 0.56$ & $10.35\pm0.21$ & $9.45\pm0.20$ & $5.53\pm0.17$ & $3.52\pm0.13$ \\
      50-60\% & $47.79\pm 0.42$ & $48.29\pm 0.42$ & $6.00\pm0.16$ & $5.49\pm0.15$ & $3.24\pm0.13$ & $2.18\pm0.10$ \\
      60-70\% & $26.91\pm 0.31$ & $27.20\pm 0.31$ & $3.34\pm0.12$ & $3.09\pm0.11$ & $1.78\pm0.09$ & $1.28\pm0.08$ \\
      70-80\% & $13.76\pm 0.22$ & $13.87\pm 0.22$ & $1.70\pm0.08$ & $1.58\pm0.08$ & $0.89\pm0.06$ & $0.67\pm0.05$ \\
\hline \hline
\end{tabular}
\end{table}

\vspace{1\baselineskip}

{\subsection{Particle ratios}}

\begin{figure}[htbp] \centering
\setlength{\abovecaptionskip}{-1cm}
\includegraphics[angle=0,width=17.6cm]{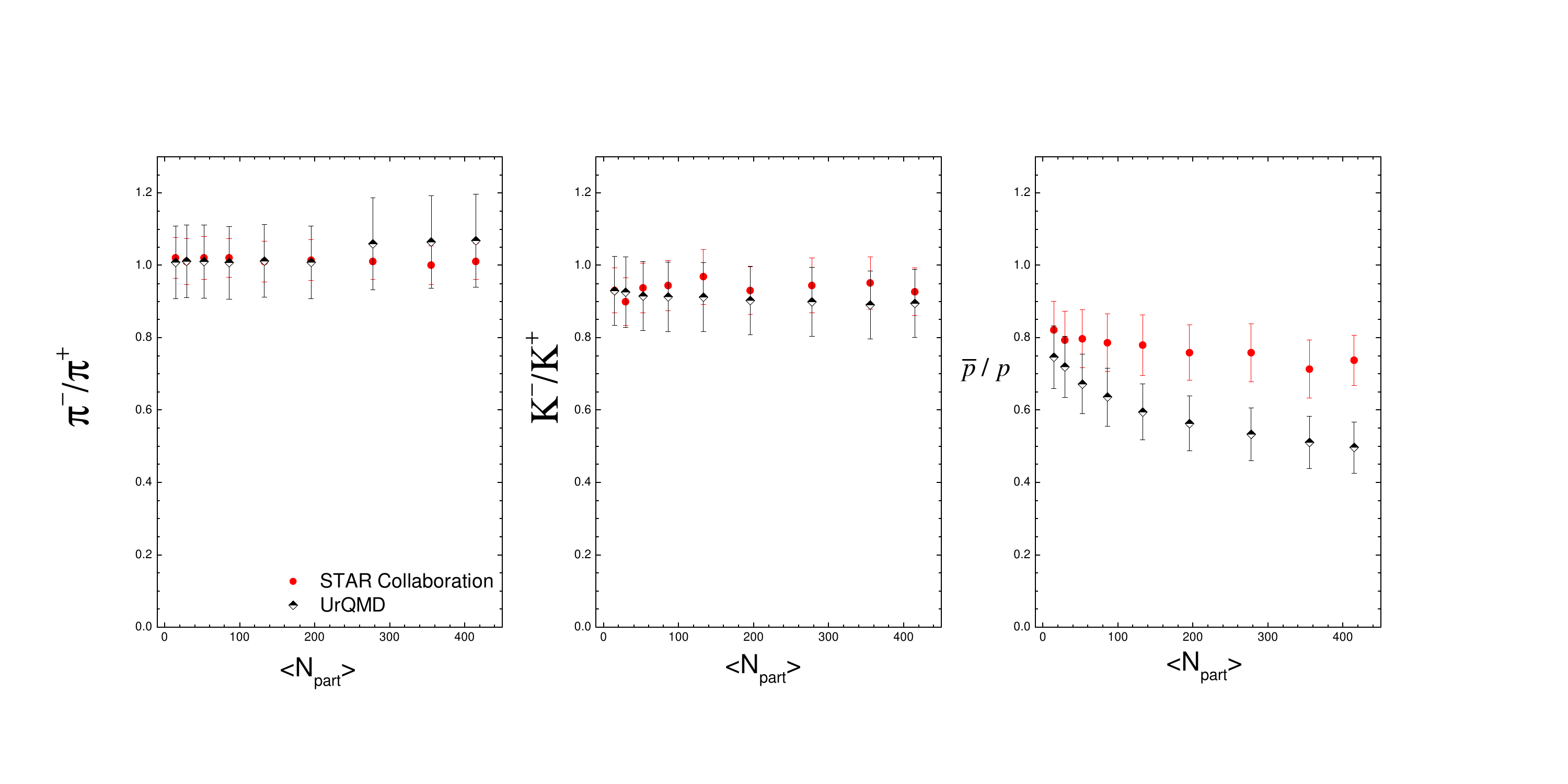}
\caption{$\pi ^{-} / \pi ^{+}$, $K ^{-} / K ^{+}$ and $\overline{p} / p$ as a function of $\left \langle N_{part}  \right \rangle$ at mid-rapidity ($\left | y \right | < 0.1$) for U+U collisions at $\sqrt{s_{NN}}$=193 GeV. The experimental data of the STAR Collaboration \upcite{Abdallah25} used in this analysis is shown as red solid circles. The black diamonds are the calculation using UrQMD model.} \label{fig6}
\end{figure}

Fig.\ \ref{fig6} illustrates the ratios of $\pi ^{-} / \pi ^{+}$, $K ^{-} / K ^{+}$, and $\overline{p} / p$ as a function of $\left \langle N_{part}  \right \rangle$ at midrapidity ($\left | y \right | < 0.1$) for U+U collisions at $\sqrt{s_{NN}} = 193$ GeV. The black diamonds represent results from UrQMD calculations, while the red solid circles denote experimental data as reported in Ref. [25]. It is observed that the experimental findings align well within the margin of error. For both $\pi$ and $k$, the yield ratios are nearly identical and remain approximately constant across different conditions. This indicates that positive and negative mesons are produced in pairs. In contrast, for the ratio of $\bar{p}/p$, there is a slight downward trend observed from peripheral to central collisions. This indicates that the more central the collision, the stronger the blocking effect on protons. Considering that the UrQMD model does not explicitly include the quark-gluon plasma (QGP) phase transition, it may underestimate the corrective effect of the collective flow effect between partons in the central collision on the antiproton/proton yield ratio.

\vspace{1\baselineskip}

{\section{Summary and Outlook}}

In this study, we systematically investigated the transverse momentum ($p_{T}$) spectra of $\pi^{\pm}$, K$^{\pm}$, and p(${\bar{p}}$) in mid-rapidity ($|y| < 0.1$) for nine centrality classes ($0–80\%$) in  collisions at $\sqrt{s_{NN}} = 193$ GeV using the Ultra-Relativistic Quantum Molecular Dynamics (UrQMD) model. We also analyzed centrality-dependent observables, including the average transverse momentum $ \langle p_{T} \rangle$, particle yields ($dN/dy$), and particle ratios, comparing our results with experimental data from the STAR Collaboration \upcite{Abdallah25}. Our analysis reveals that the collision dynamics are highly sensitive to the prolate deformation of the uranium nucleus. This non-spherical geometry leads to significant variations in the initial compression and lifetime of the high-density matter depending on the relative orientation of the colliding nuclei. Consequently, the description of particle production requires different theoretical approaches in distinct kinematic regions \upcite{wu40}. The cascade mode provides a robust description of the experimental data in the low-$p_{T}$ region ($p_{T} < 1.2 GeV/c$), where hadronic rescattering and kinetic equilibrium dominate. In contrast, the SM-EoS model, which incorporates a soft momentum-dependent equation of state, shows better agreement with data in the intermediate-to-high $p_{T}$ region ($1.2 GeV/c < p_{T} < 2.0 GeV/c$), highlighting the importance of mean-field effects in this regime.

We note that discrepancies between the SM-EoS calculations and experimental data persist, particularly in peripheral collisions, which may be attributed to the simplified treatment of the equation of state in the high-density regime. Furthermore, since the UrQMD model is a hadronic transport framework, it does not explicitly include partonic degrees of freedom or a QGP phase. This limitation contributes to the deviations observed at higher $p_{T}$. Additionally, the antiproton-to-proton yield ratios approaching unity in mid-rapidity indicate a high degree of matter-antimatter symmetry, suggesting that particle production in this region is consistent with chemical equilibrium or dominant pair-production mechanisms. Future work will focus on enhancing the model by integrating partonic degrees of freedom or coupling UrQMD with hydrodynamic simulations to better describe the high-$p_{T}$ spectra and the QCD phase transition dynamics in deformed nuclear collisions.

\vspace{1\baselineskip}

{\section*{Acknowledgements}}

We are grateful to the C3S2 computing center in Huzhou University for calculation support. This study used computational resources provided by Institute of Marine Drugs, Guangxi University of Chinese Medicine and the special Fund for Hundred Talents Program for Universities in Guangxi (Gui2019-71). The authors acknowledge the Beijing Super Cloud Computing Center (BSCC) for providing HPC resources that have contributed to the research results reported within this paper. URL: http://www.blsc.cn/ This work was supported by the Fund for Less Developed Regions of the National Natural Science Foundation of China under Grant No.12365017, the Natural Science Foundation of Guangxi Zhuangzu Autonomous Region of China under Grant No. 2021GXNSFAA196052, the Introduction of Doctoral Starting Funds of Scientific Research of Guangxi University of Chinese Medicine under Grant No.2018BS024, and the Open Project of Guangxi Key Laboratory of Nuclear Physics and Nuclear Technology, No. NLK2020-03.

{\section*{Conflict of Interests}}

The authors declare that there is no conflict of interests regarding the publication of this paper.

\end{CJK}
\end{document}